# EVLearn: Extending the CityLearn Framework with Electric Vehicle Simulation

Tiago Fonseca*, Luis Ferreira*, Bernardo Cabral, Ricardo Severino
INESC-TEC/Polytechnic of Porto - School of Engineering
Porto, Portugal
{calof*, llf*, bemac, sev}@isep.ipp.pt

Kingsley Nweye, Dipanjan Ghose, Zoltan Nagy
The University of Texas at Austin
Austin, 78712, Texas, USA
{nweye, dipanjan02, nagy}@utexas.edu

*Abstract*— Intelligent energy management strategies, such as Vehicle-to-Grid (V2G) and Grid-to-Vehicle (G2V) emerge as a potential solution to the Electric Vehicles' (EVs) integration into the energy grid. These strategies promise enhanced grid resilience and economic benefits for both vehicle owners and grid operators. Despite the announced prospective, the adoption of these strategies is still hindered by an array of operational problems. Key among these is the lack of a simulation platform that allows to validate and refine V2G and G2V strategies. Including the development, training, and testing in the context of Energy Communities (ECs) incorporating multiple flexible energy assets. Addressing this gap, first we introduce the EVLearn, a simulation module for researching in both V2G and G2V energy management strategies, that models EVs, their charging infrastructure and associated energy flexibility dynamics; second, this paper integrates EVLearn with the existing CityLearn framework, providing V2G and G2V simulation capabilities into the study of broader energy management strategies. Results validated EVLearn and its integration into CityLearn, where the impact of these strategies is highlighted through a comparative simulation scenario.

## I. INTRODUCTION

Renewable Energy Sources (RES) and Electric Vehicles (EVs) are emerging as pivotal players in the shift towards a low-carbon economy [1]. Distributed Energy Resources (DERs), such as small-scale wind and solar production, are facilitating the involvement of consumers in the energy transition by increasing adoption of rooftop solar [2].

However, fully integrating EVs, RES and DERs into the current energy grid infrastructure presents a range of infrastructural, control, and technological challenges [3], [4]. One example is the mismatch between peak EV charging and RES production, leading to an imbalance between the demand for renewable energy and its supply which mainly occurs during the sunlight hours. As documented with the duck curve [5], it implies costly mitigation efforts, and could potentially impede the environmental benefits of these innovations or even the widespread use of EVs and solar power [6]. Therefore, researchers in [7] point to the need to embark on a significant transformation in how we generate and consume energy, towards intelligent energy management strategies.

### A. Energy Management

Intelligent energy management provides solutions to the mentioned problems by controlling energy resources more efficiently [8]. Central to these solutions are Demand Response (DR) [9], energy flexibility optimization approaches [10], and Energy Community (EC) management [11], which strive to balance energy supply and demand, reducing costs, and enhancing grid stability.

Also, at the heart of this evolving landscape of energy management is Vehicle-to-Grid (V2G) technology, a concept that seeks to integrate EVs into the broader power grid management given their load shifting flexibility [12]. V2G leverages parked EVs into mobile energy storage units that not only draw power from the grid for charging but also can feed electricity back into the grid during periods of high demand [13]. This dual function will position EVs as significant players in power management, contributing to grid reliability, while potentially offering financial benefits for their owners [14]. In parallel, simpler strategies, such as smart charging Grid-to-Vehicle (G2V), which shifts energy loads to during low demand or high renewable generation [15].

Such management approaches, hereby called Energy Management Systems (EMS), require advanced algorithms that optimize control given factors like grid conditions, energy prices, vehicle usage patterns, prosumer flexibility, battery health, among others [16]. Algorithms can range from Machine Learning (ML), expert tuned rule-based control, to Model Predictive Control (MPC), meta-heuristics and optimal control algorithms. RL, in particular, has gained significant popularity in the research community [17].

### B. The need for simulation

Given the critical nature of energy systems, deploying untested EMS approaches directly in the real-world is impractical and risky [18]. As such, simulators provide a safe and controlled environment where these algorithms can be rigorously trained and refined without endangering the grid but also potentially compromise occupant comfort and preferences, such as risking an uncharged electric vehicle (EV) at the time of departure. Simulators allow for the modeling of various scenarios, including rare and extreme conditions, ensuring the algorithms are robust and reliable [19]. The role of simulators for energy applications is extended when using and training RL approaches which depend upon trial-and-error interactions with the grid to learn optimal actions [17].

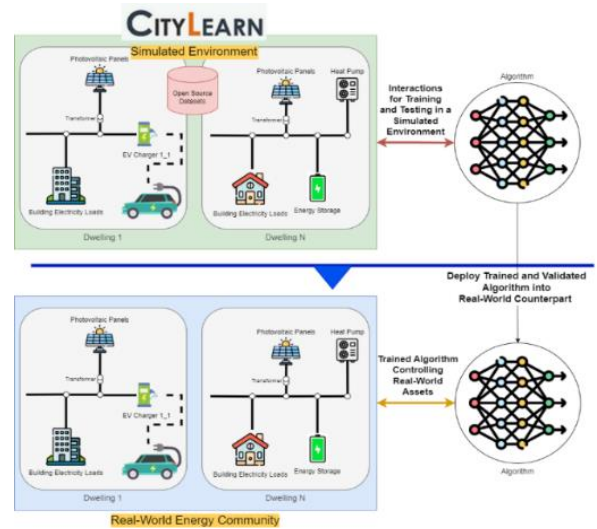

*Figure 1- Simulators role to the real-world implementation of EMSs*

Therefore, a standardized and realistic simulation environment is vital for benchmarking and comparing EMSs [20]. It enables developers to create a virtualized realistic environment where different algorithms can be evaluated, understanding strengths and nuances of each under identical conditions. This not only facilitates the selection of the most effective solutions but also fosters collaboration and progress within the research community. By sharing results and methodologies in a common environment, researchers can build on each other's work, accelerating innovation and driving the field forward (Fig. 1.).

*C. Gaps and Contributions*

Despite the significant progress made in both EMSs and on the standardizations of testing environments, to the best of the authors knowledge there is no standardized simulation framework which integrates EV energy management with the diversity of control of other energy community assets, such as solar panels and heat pumps, into the same standardized simulation framework. Although one can argue that this is partially due to the novelty of the V2G concept, another justification, and perhaps more concerning, is the limited focus of current research in the interactions of energy assets considered. Indeed, most research focuses on the optimization of one kind of asset at a time [8]. As pointed by [6], the best strategy for energy management cannot only consider one type of asset or a single building alone; on the contrary, it should consider the diversity and heterogeneity of multiple assets, aggregating their energy flexibility for better and most effective management.

Given the identified research gap, this paper focuses on the integration of these two areas, by effectively creating an EV simulation extension module, the EVLearn, which is designed to integrate to an established simulator for EMSs which already models other energy assets, the CityLearn framework [21]. EVLearn extension modulates characteristics of charging stations, charging behaviors, load shifting flexibility, EV batteries, and factors in pre-simulated vehicle usage patterns. The paper also describes the work done to integrate EVLearn into the CityLearn framework, which is detailed in Chapter II. This work's necessity is also highlighted by previous research work that have mentioned the need to expand the CityLearn framework with EVs simulation [20], [21]. The paper's contributions can be resumed as follows:

- Design and implementation of EVLearn, an EV energy management simulation module with charging, load shifting flexibility and vehicle usage modulation.
- Integration of EVLearn within an established simulator, the CityLearn framework, providing a complete testbed (with a broad scope of energy assets) for researchers developing energy management algorithms.
- Demonstration of the validity and integration of EVLearn with a created simulation scenario.

*D. Paper Outline*

The remainder of the paper can be resumed as follows. In Chapter II, other approaches into simulation platforms are explored, and the CityLearn framework is detailed. Chapter III describes the developed EVLearn. Chapter IV details the integration of EVLearn into CityLearn. The experimental setting for testing the developed work are presented in Chapter V. Chapter VI showcases the preliminary results of EVLearn, while Chapter VII concludes and outlines future directions.

## II. RELATED WORK

Current tools, such as Energym [22] and BOPTEST [23], offer high-fidelity energy models for control algorithm benchmarking. These emulators leverage advanced simulation engines like EnergyPlus [24] and Modelica [25], providing a robust foundation for modeling thermodynamics and control systems in built environments. These are primarily designed for system-level or building-level simulations and do not adequately cater to the complex interactions or the scalability required for effective V2G and G2V simulation.

Similarly, broader energy system simulation tools such as the System Advisory Model (SAM) by NREL [26] and GridSim [27] incorporate models for solar PV systems, energy storage, and scheduling. Low-level simulation frameworks like Pandapower [28] arevaluable for DC optimal power flow calculations and distribution network simulations. However, all these fall short in addressing the specific challenges posed by V2G and G2V technology, such as including the definition of EV's energy flexibility (i.e., when will the owner of the car have it parked so that the vehicle can participate in such strategies to provide grid services).

Besides the simulation frameworks previously presented, CityLearn [21] is a standardization tool for facilitating the implementation, testing, and benchmarking of centralized and decentralized RL, MPC, Rule-Based Controllers (RBC) algorithms for DR, load shifting and urban energy management. Its environment offers a wide range of parameters, including various building types, HVAC systems, and weather conditions across different climatic zones. This is particularly relevant in contexts such as ECs or energy districts. Despite modelling this assets and creating a solid testbed for algorithms, CityLearn does not model EVs, their charging infrastructure and the simulation of energy flexibility needed for V2G and G2V. Given the projected increase in EV adoption and its associated impact on energy demand, simulating EVs within the CityLearn environment could provide valuable insights and offer a more comprehensive training, testing and benchmarking ground for RL algorithms, bringing simulations closer to real-world scenarios.

## III. EVLEARN DESIGN

One of the fundamental real-world dynamics introduced by the integration of EVs into energy management strategies is the inherent variability of their connection and availability status. Unlike fixed components, such as buildings' heat pumps and stationary batteries already modeled into CityLearn, EVs can transition between plugged and unplugged (travelling) states as their primary functionis to engage in regular transportation of people and goods. This mobility brings forth a degree of unpredictability and adds a layer of complexity to the optimization strategies, and thereby needs to be modelled into the simulation (Section III.E).

In EVLearn environment, energy management of EVs is designed to allow the simulation of three distinct dynamics: (i) V2G; (ii) Grid-to-Vehicle (G2V); (iii) No Control (i.e., where EVs act as a load without any possible control over their charging). For that purpose, EVLearn has three fundamental parts: the Electric Vehicle Charger's (EVCs), which serve as a connection between a building and an EV (Section III.A.); the model of EV itself, which acts as a flexible DER (Section III.B.); and, a pre-simulated dataset, which dictates the plug in/out energy flexibility routine for each EV and introduces formulation for energy flexibility (Sections III.C and III.D.).

## A. Electric Vehicle Charger Model

The EVC represents the physical infrastructure required for transferring power between these entities, either to charge up the EV battery or to return stored energy back to the grid. As better detailed in Section III.C., this is the component that the EMS (centralized or decentralized) will have control over.

### 1) Modelling One or Multiple Chargers per Building

Just like a house, an office or any building can have multiple installed chargers in the real-world, a single building in the simulation can have more than one charger simulated. In EVLearn, a simulated charger is modeled to replicate a real-world plug of a charger, so more than one may be necessary at each building. This is useful for simulating the need to manage multiple EVs simultaneously in a single dwelling, either as a home with several EVs, an office building with a set of chargers available to their workers, or even simulating a public charging hub with multiple stalls available. Each charger is assigned a unique identifier $EVC_{b\_n\_p}$ (where $EVC$ stands for Electric Vehicle Charger, $b$ stands for the Building where the charger is inserted in the simulation, $n$ for the charger Number within the building and $p$ stands as the number of the Plug of that charger). This will facilitate the appropriate linkage between the EVC and the EVs during the simulation.

### 2) Modelling Energy Management Control

When an EV is plugged in, the electricity consumption of the electric vehicle charger $E_t^{EVC_{b\_n\_p}}$ is a function of the control action decided by the control algorithm, $a_{t-1}^{EVC_{b\_n\_p}}$, where $a \in [-1, 1]$ for V2G and $a \in [0, 1]$ for G2V and No-Control dynamics. The action $a_{t-1}^{EVC_{b\_n\_p}}$ denotes the proportion of charger's nominal power $p^{EVC_{b\_n\_p}, \text{ nominal, charging}}$ when action $a_{t-1}^{EVC_{b\_n\_p}} > 0$ and the proportion of charger's nominal power for discharging $p^{EVC_{b\_n\_p}, \text{ nominal, discharging}}$ when $a_{t-1}^{EVC_{b\_n\_p}} < 0$ (Eq. 1.). The supplied charging energy from the charger $Q_t^{EVC_n}$ (Eq. 2.) is the product of the electricity consumption $E_t^{EVC_{b\_n\_p}}$ and technical efficiency $\eta^{EVC_{b\_n\_p}, \text{ technical}}$.

$$E_t^{EVC_{b\_n\_p}} = \begin{cases} a_{t-1}^{EVC_{b\_n\_p}} \times p^{EVC_{b\_n\_p}, \text{ nominal, charging}}, & a_{t-1}^{EVC_{b\_n\_p}} \geq 0 \\ a_{t-1}^{EVC_{b\_n\_p}} \times p^{EVC_{b\_n\_p}, \text{ nominal, discharging}}, & a_{t-1}^{EVC_{b\_n\_p}} < 0 \end{cases} \quad Eq.\ 1$$

$$Q_t^{EVC_{b\_n\_p}} = \eta^{EVC_{b\_n\_p}, \text{ technical}} \times E_t^{EVC_{b\_n\_p}} \quad Eq.\ 2$$

## B. Electric Vehicle Model

The EV model replicates the real-world operational attributes and practical constraints of EVs within a simulation environment, specifically those acting as a significant energy flexibility factor within the system, e.g., participating in one of the defined energy management dynamics. EVs in the simulation can be connected to a EV Charger (Section III.B.) and consume energy (in all dynamics) and discharge back to the grid (in V2G dynamics). EVs will connect and disconnect from the chargers as per the dynamic pre-simulated file described at Section III.D.

At the heart of the EV model is its battery model, derived from the formulation of CityLearn framework's Stationary Battery. The EV Battery Energy Storage System (EV BESS) has a time-dependent capacity, $C_t^{EVBESS}$, which reflects the battery degradation over time (i.e., the maximum capacity value decreases through time). It also incorporates a round-trip efficiency, $\eta^{EVBESS, round-trip}$, which accounts for losses regarding the exchange of energy (except the efficiency of the charger which is already accounted at Eq. 2.). Moreover, the EVBESS defines a maximum input and output power, $P_t^{EVBESS}$. Note that this value already factors in (Eq. 3.) the product of the nominal power, $P^{EVBESS,nominal}$ and SoC-power dependent function $f(SoC_{t-1}^{EVBESS})$. This function determines the proportion of nominal power made available at any $SoC_{t-1}^{EVBESS}$.

$$P_t^{EVBESS} = P^{EVBESS,nominal} \times f(SoC_{t-1}^{EVBESS}) \quad Eq.\ 3$$

The stored energy, at any time step, $Q_t^{EVBESS}$, is a piecewise function (Eq. 4.). This function is driven by the energy supplied by the charger $Q_t^{EVC_{b\_n\_p}}$ to where the EV is connected. $Q_t^{EVC_{b\_n\_p}}$ determines the energy amount the battery will receive for charging (when > 0) or discharging (when < 0).

$$Q_t^{EVBESS} = \begin{cases} Eq.\ 5, & Q_t^{EVC_{b\_n\_p}} \geq 0 \\ Eq.\ 6, & Q_t^{EVC_{b\_n\_p}} < 0 \end{cases} \quad Eq.\ 4$$

The energy stored in an EVBESS after charging is determined by Eq. 5. This equation states that the stored energy $Q_t^{EVBESS}$ at any given time $t$ is limited to the lesser of two values: the system's capacity ($C_t^{EVBESS}$), as the battery cannot charge up more than its capacity, and an expected energy value. The expected energy value itself is calculated from two main components.

Firstly, we have the initial energy remaining in the battery, $Q_{t-1}^{EVBESS}$, from the previous period, adjusted for thermal losses. This is represented by $Q_{t-1}^{EVBESS} \times (1 - \theta^{EVBESS})$, where $\theta^{EVBESS}$ is the thermal loss coefficient. Essentially, it accounts for energy lost due to heat, providing us with the actual energy available at the start of the current charging period.

Secondly, the expected energy includes the amount of energy to be added during this period. This is calculated by taking the minimum of two quantities: the energy coming from the control action applied to the EVC to where the EV is connected, $Q_t^{EVC_{b\_n\_p}}$, and the maximum power the battery can accept or deliver at that state of charge (SoC), noted as $P_t^{EVBESS}$. This value is then multiplied by the system's round-trip efficiency, $\eta^{EVBESS, round-trip}$, which reflects the efficiency losses in storing and retrieving energy.

$$\begin{aligned} Q_t^{EVBESS,+} = \min \Big( &C_t^{EVBESS}, Q_{t-1}^{EVBESS} \times (1 - \theta^{EVBESS}) \\ &+ \min \big( Q_t^{EVC_{b\_n\_p}}, P_t^{EVBESS} \big) \\ &\times \eta^{EVBESS, round-trip} \Big) \end{aligned} \quad Eq.\ 5$$

Contrarily, the stored energy after discharging the EV BESS is defined by Eq. 6. This equation captures the total energy remaining after discharging activities are accounted for, which involves two primary calculations. First, we assess the initial energy in the battery but subtract any energy losses, providing a baseline of available energy before any discharge occurs, $Q_{t-1}^{EVBESS} \times (1 - \theta^{EVBESS})$.

Second, the equation considers the energy that has been drawn out for use $Q_t^{EVC_{b\_n\_p}}$, by the charger to which the EV is connected to. This energy is limited by the system's capacity to output power (denoted as $-P_t^{EVBESS}$) and the round-trip efficiency, $\eta^{EVBESS, round-trip}$. The stored energy is limited to a depth-of-discharge (DoD) such that the BESS is never completely drained for DoD > 0.

$$Q_t^{EVBESS,-} = \min \Big( C_0^{EVBESS} \times DoD^{EVBESS}, Q_{t-1}^{EVBESS} \\ \times (1 - \theta^{EVBESS}) \\ + \min \big(Q_t^{EVC_{b\_n\_p}}, -P_t^{EVBESS}\big) \\ \div \eta^{EVBESS,round-trip} \Big) \quad Eq.\ 6$$

The EVBESS SoC at any given time step $t$, is defined by equation Eq. 7 as a function of the stored energy and capacity before any degradation begins.

$$SoC_t^{EVBESS} = \frac{Q_t^{EVBESS}}{C_0^{EVBESS}} \quad Eq.\ 7$$

### C. EVLearn Simulation Dynamics

After defining the EV model (Section III.B.) and the EVC model (Section III.A.) as the base of EVLearn, this Section overviews the design decisions that connect these models into the simulation of the three distinct dynamics: (i) V2G; (ii) Grid-to-Vehicle (G2V); (iii) No Control.

#### 1) Controlability

In the simulation dynamics EVLearn, we established that the control over EVs charging is exercised exclusively when the vehicles are plugged into an EVC. Specifically, EMSs (either centralized or decentralized) regulate the chargers delivery, which in turn charges/discharges the connected EVs.

As such, EMSs will not have an influence when an EV arrives or leaves (plugs in or out from a charger); these events are dictated by the real-life schedules and habits of the vehicle owners, which are pre-simulated according to the concept of energy flexibility (Section III.C.2)). Thus, the role of EMSs is not to direct vehicle owners on when to connect or disconnect their EVs. Instead, their function is to optimally manage charging during the times the vehicles are available (connected to a charger within a building), treating them as DERs within the system.

#### 2) Energy Flexibility

To simulate these dynamics, the concept of energy flexibility is utilized, adapted from the standardized FlexOffer (FO) model, introduced by [29]. A FO for an EV expresses a tuple with: the pre-determined conditions of the earliest start time $t_{es}$ (when the EV connects to an EVC), the latest start time $t_{ls}$ (when the EV disconnects from an EVC), a SoC required at departure $SoC_{departure}$ and an energy profile (a list that contains a sequence of slices $s$ that represent the energy profile of charging/discharging) comprehended within $t_{es}$ and $t_{ls}$ (Eq. 8.). Each slice $s_{tn}$, of duration 1-time unit, $t$, is in an energy range between and , usually represented in kWh, which can be positive if the device consumes energy or negative if the device returns energy back into the grid.

$$FO_n^{EV} = \big([t_{es}, t_{ls}, SoC_{departure}], [s_{t1}, s_{t2}, \ldots, s_{tn}]\big) \quad Eq.\ 8$$

Considering the operational flexibility of an EV in relation to the power grid, the energy profile can be defined within the FlexOffer model (Eq. 9.)

$$s_{tn} = [p_{tn}^{min}, p_{tn}^{max}] \quad Eq.\ 9$$

where $s_{tn}$ represents a slice of time within the FO, and $p_{tn}^{min}$ and $p_{tn}^{max}$ delineate the minimum and maximum power, respectively, for that time slice. The maximum, $p_{tn}^{max}$, corresponds to the lowest of either the charger's supplied energy $p^{EVC_{b\_n\_p},\ nominal,\ charging}$ or the EV's battery maximum charging power, $P_t^{EVBESS,charging}$ at time $t_n$ (Eq. 10.).

$$p_{tn}^{max} = \min\big(p^{EVC_{b\_n\_p},\ nominal,\ charging}, P_t^{EVBESS,charging}\big) \quad Eq.\ 10$$

The determination of $p_{tn}^{min}$ (Eq. 11.) varies based on the operational mode. For V2G dynamics, the EV can return energy to the grid, facilitating energy distribution during peak demand or other strategic periods. As such, the minimum power, $p_{tn}^{min}$ is defined negatively to represent energy discharge. For G2V and No-Control Dynamics, as the $p^{EVC_{b\_n\_p},\ nominal,\ discharging}$ of the charger is set to 0, as there is no energy return to the grid.

$$p_{tn}^{min} = -\min\big(p^{EVC_{b_{n_p}},\ nominal,\ discharging}, P_t^{EVBESS,discharging}\big) \quad Eq.\ 11$$

Translating from the mathematical definition, an example of a FO within the EVLearn is: "EV is connected to EVC, I want to leave at 8 AM with its battery charged to 80 per cent SoC." In V2G and G2V dynamics, finding the most optimized control for energy use on slices, within power delivery boundaries, will be the work of the EMS. In special cases, $t_{ls}$ can be undefined (i.e., when an EV owner does not know when will leave). In these cases, the EMS will consider it as a stationary battery until further notice of a predicted departure.

### D. Pre-Simulated EV Energy Flexibility Dataset

The EVLearn module has not been designed to directly simulate the commuting patterns and daily routines of EVs, nor does it inherently model the energy consumption, charging, or disconnection behaviors associated with EVs traveling between locations (FOs defined at Section III.C.). Instead, these aspects are addressed by incorporating pre-simulated datasets into the simulation.

Motivated by the absence of suitable existing datasets on EV energy flexibility [30], [31], and to enhance the realism of the simulation, a series of Python scripts have been developed to generate data through a synthetic process. These scripts are capable of simulating EV commuting behaviors, energy expenditure during travel and other energy flexibility dynamics based on a range of statistical parameters that reflect real-world variability in vehicle connection statuses and owner's mobility. Each generated dataset, specific to an individual EV, is a time-series that includes, at least, information on arrival and departure times, the needed SoC at departure and the EVC id to which the EV is connecting. This information is necessary to generate FOs for EV energy management within the EVLearn simulation framework.

Besides the aforementioned data, the dataset can also provide forecast data to EVLearn. This will provide the EMS with the ability to anticipate and manage the energy demand arising from EVs planning to return home and charge by providing an expected arrival time and a forecasted SoC at arrival. In real-world implementations this can be achieved by allowing EVs to signal charging intentions through navigation systems, enabling the simulation to prepare for upcoming charging events. Such a mechanism facilitates the modeling of EMSs that take into account future charging operations.

The synthetic generation script involves two distinct modes of operation: one simulating the charging behavior and household routines, and the other focusing on the behavior of EVs in relation to workplace charging and work schedules. The integration of this tailored dataset allows for a more comprehensive and accurate representation of EV behavior and its implications for EMSs for different real world settings.

In the household generation mode the data depends on the setting of different parameters, such as at what time the EV typically leaves home and its typical statistical variation, also for the time of arrival home. From our observation of different datasets and self-experience it is also possible to add a certain probability for major changes in this routine. Regarding weekends, certain habits like going out in the morning for a walk or shopping, or going out in the afternoon for hobbies, among other things, are also considered. For the workplace mode it is possible to define the probabilities of encountering heavy traffic during commuting, which basically translates delays on the morning arrival to the office charger. Another factor used in this mode is the distance from home and the associated reduction on the SoC. For a more detailed description of the parameters for setting the synthetic data generation refer to REF.

The output of the referred scripts is a set of files, one for each EV that exists in the scenario. The files are structured as a series of arrays, each representing a distinct aspect of the EVs' operation at specific day and time where:

- **Timestamp**: a time series of the months and hours, ranging from 1 (January) to 12 (December) and from 1 to 24, respectively. When integrating with CityLearn, this should match the time of the simulation of CityLearn.
- **EV_State**: depicts the state of the EV, denoting whether it is plugged in and ready to charge (represented as 1) or incoming to a charger, not connected (represented as 2), or in transit, not connected (represented as 3).
- **Charger**: specifies the charger where the EV is or will be plugged in during its next plugged in and ready to charge state. It can contain 'nan' if no destination charger is specified, or the charger ID. This information is not directly used by the simulator, but will be used by logic inserted within the `next_time_step()` function, described at IV.B.
- **Est_Departure:_Time**: provides the number of time steps expected until the vehicle departs. This data is only available when the EV is in the ready to charge state.
- **Req_SOC_Departure**: provides the estimated SoC that the EV requires at departure time.
- **Est_Arrival_Time**: number of time steps expected until the vehicle arrives at a charger.
- **Est_SoC_Arrival**: estimated SoC percentage for the EV at arrival time.

An example dataset is provided with the EVLearn integration into CityLearn's main code repository. This synthetically generated dataset provides data on 12 EVs, 8 of which associated with households in the district and 4 of which associated with one office building. The behavior and energy flexibility of the EVs associated only to the Office building are different from those associated with households, as they charge during the day and compete for the only two available chargers within the office.

## IV. INTEGRATION OF EVLEARN INTO CITYLEARN

Integrating EVLearn into the CityLearn framework mandated a critical emphasis on backward compatibility. This requirement is pivotal, ensuring that existing EMSs and configurations within CityLearn remain functional and undisrupted by the new extension. Only this way, the EVlearn integration can be merged into the main code repository and enrich the research community's resources. In addition to backwards compatibility, adherence to the established conventions of CityLearn, such as definitions of observations and actions logic, schema definitions, and objectives, is essential. This section details the undertaken development and design decisions to achieve the following objectives:

- Seamless integration of EVLearn into the architecture and loop structure of CityLearn, negating the need for supplementary code. This extension maintains ease of installation and use, necessitating minimal additional dependencies and minor schema modifications for EV modeling (Sections IV.A. and IV.B.).
- Energy flexibility in the form of FOs (according to a pre-simulated file) is translated into the observations and actions of CityLearn (Section IV.C.).
- Introduction of a new, customizable reward function tailored to the specific requirements of EV energy management dynamics (V2G, G2V), enhancing the framework's adaptability to research of EV-related EMSs scenarios (Section IV.D.).

*A. Design*

EVLearn uses the underlying structure of CityLearn, Rooted in the OpenAI Gym standard [32]. As such, the *Environment* class inherited from Gym as its core class for implementation. The domain model of the existing simulation platform included buildings, electric heaters, heat pumps, thermal storage systems, stationary batteries, and PV systems (depicted in grey Fig. 2.).

Following the requirements analysis, the primary and most important modification to the domain model were the incorporation of EVs, EVC and associated environment logic (dynamics Section III.C.). To this end, two distinct classes were introduced: `EV` class and `Charger` class (depicted in light green in Fig. 2.). These translate into CityLearn the modelling presented during Section III.

Note that as specified during the dynamics, EVs, may, or may not be connected to a EVC. To this end, `EV` class inherits directly from the `CityLearn Env` class as its simulation is independent from the buildings and given according to the pre-simulated file (Section III.D.), which dictates the plug in/out energy flexibility routine for each EV and introduces formulation for energy flexibility. The EVC are part of a building defined in the CityLearn, and inherit properties of the standalone electrical devices.

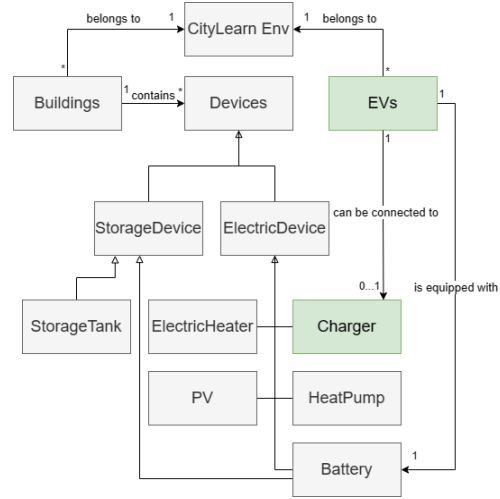

*Figure 2 - City Learn's Domain Model*

One or more EVC can be defined per building (according to the nomenclature defined in Section III.B.). EVC act as the connection point between the Building and the EVs. The main point of control continues to be a per building control, as EMSs (centralized or distributed) will only control building's EVCs, and not EVs directly.

*B. Step Function Extension*

Updates to the fundamental loop functions, such as `reset` and `step` are also part of the work done to integrate EVLearn into CityLearn. In the simulation environment the `step` function plays a pivotal role in orchestrating the dynamic of EvLearn, as it translates the logic from the pre-simulated EV energy flexibility file into the system. Its implementation updates the function with the same name in CityLearn. Fig. 3. presents a simplified sequence diagram of these functions and how the EVLearn dynamic is integrated (in light green).

Upon initiation, the `step` function starts by receiving and processing a series of actions. The actions corresponding to the management of dwellings' energy assets (existing previously in the City Learn framework) are applied first. Then, for each charger $EVC_{b,n,p}$, the action $a_{t-1}^{EVC_{b\_n\_p}}$, is applied to charger $EVC_{b\_n\_p}$ (`update_EV_storage(action)`) and then to the connected EV (`charge_discharge_battery(Q)`).

Finally, the environment advances into the next time step, *t+1*, and before returning the new observations for *t+1*, the `advance_evs()` translates the energy flexibility in the pre-simulated files into CityLearn. This includes connecting or disconnecting EVs to and from EVCs based on data sourced from each the associated dataset file.

The pre-simulated dataset states a SoC at arrival (Section III.D.), however, to add to the stochastic of events, when connecting an EV to an EVC, the value is randomized within a normal distribution. In this way, the simulation of the EV continues to evolve, imitating real-world conditions where vehicles have varying states of connection and mobility that can be failed to be predicted.

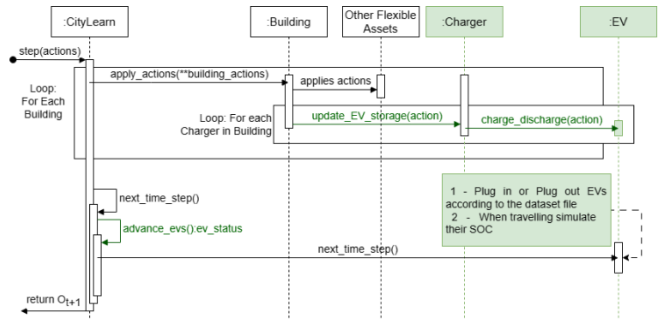

*Figure 3 -Sequence diagram of the updated step function*

*C. Observations and Actions*

Observations and actions are crucial for the RL algorithms (or any type of EMSs as they act as the environment interaction: input and output). Observations provide the EMS with information about the environment, its current state, and the various factors that might influence the agent's decisions, acting like "sensors" in real world. Actions, on the other hand, are the means through which the agent can control the environment. There are over 50 pre-defined observable continuous states in CityLearn. To maintain the integrity of CityLearn the original observations and actions remain intact.

*Table 1 – New Observations*

| Name | Description | Source |
|---|---|---|
| $O^{EVCb,n,p\_S}_t$ | Charger State | ElectricVehicle_id.csv - State |
| $O^{EVCb,n,p\_EDT}_t$ | Charger Connected EV Estimated Departure Time | electric_vehicle_id.csv – Est_Departure_Time |
| $O^{EVCb,n,p\_SoC\_D}_t$ | Charger Connected EV Required SoC Departure | electric_vehicle_id.csv – Req_SOC_Departure |
| $O^{EVCb,n,p\_EAT}_t$ | Charger Incoming EV Estimated Arrival Time | electric_vehicle_id.csv – Est_Arrival_Time |
| $O^{EVCb,n,p\_SoC\_A}_t$ | Charger Incoming EV Estimated SOC Arrival | electric_vehicle_id.csv – Est_SoC_Arrival |
| $O^{EVCb,n,p\_SoC}_t$ | Charger Connected EV SoC | calculated |

In addition to these, new observations have been introduced to encompass the dynamics of EVs energy management and their energy flexibility. Observations defined for EVLearn integration into CityLearn translate the energy flexibility pre-simulated and supplied to through flat .csv files (dataset discussed in Section III.E). New observations (per EVC) are resumed at Table 1.

Note that, as previously outlined during Chapter III., the observations will only be provided from the chargers' point of view, i.e., only when a car is plugged in to a charger or incoming to a specific charger the system will be able to observe the needed information on its flexibility. When an EV is in transit, and not incoming to a charger, the EMS will not have any observations of the EV, as it also does not need that information for energy control (the EV is not participating).

This design ensures EMSs deal with the connects and disconnects logic where some of the observations are sometimes inaccessible (they cannot get data from the EVs when these are commuting) and dynamic (EVs leave with one battery SoC and arrive with a different value). The Charger State ($O^{EVCb,n,p\_S}_t$) observation translates this logic into the system and provides it to an EVC. It can have value 1 when an EV is connected to that charger, a value of 2 when an EV is incoming, or a value of 0 when no EV is connected.

Besides the connection status observation, the newly introduced observations include the EV SoC ($O^{EVCb,n,p\_SoC}_t$), which is calculated during runtime and states the actual SoC of the EV connected to a given charger (according to Eq. 7.) Moreover, analogous to the definition of Fos (Section III.C.), the EVLearn defines time and energy flexibility of EVs, such as the the Estimated Departure Time ($O^{EVCb,n,p\_EDT}_t$), and the Required SoC at Departure ($O^{EVCb,n,p\_SoC\_D}_t$) when a charger as a connected EV. Moreover, preparing for the forecasting of events (as detailed during Section III.D.), observations on the Estimated Arrival Time ($O^{EVCb,n,p\_\_EAT}_t$) and the Estimated SOC Arrival ($O^{EVCb,n,p\_SoC\_A}_t$) are given when a charger is signalling that an EV is incoming. Note that for the learning process all of these observations are used from the pre-simulated dataset file.

When the observation does not fit the current state of the Charger, for example when $EVC_{b,n,p}$ signals an incoming EV, $O^{EVCb,n,p\_\_EAT}_t$ and $O^{EVCb,n,p\_SoC\_A}_t$ are used, however $O^{EVCb,n,p\_EDT}_t$ and $O^{EVCb,n,p\_SoC\_D}_t$ are set to -1, signalling no valid observation is given. When $EVC_{b,n,p}$ states no car is connected, all other charger observations are set to -1.

Table 2 – New Action

| Name | Description |
|---|---|
| $A^{EVC_{b,n,p}\_S}_t$ | Charger Connected EV Storage |

Regarding Action, specified at Table 2, the EV Storage Action ($A^{EVC_{b,n,p}\_S}_t$), is added. Spanning the range between [-1.0, 1.0] for V2G dynamics, or the range [0, 1] in G2V and No Control Dynamics, this action specifies the percentage of the charger maximum charging or discharging capacity ($a^{EVC_{b\_n\_p}}_{t-1}$ in Eq. 1.). For example, an action of 0.5 applied to a charger with a maximum charging of 22kWh, will specify an action to charge at 11kWh. Negative values apply when the EV discharges back to the grid. Fig. 4. overviews the schematic of an EC simulation and the respective observations for each flexible asset. Note that, for clarity, some observations from the CityLearn framework are omitted in Fig. 4. to emphasize the newly introduced ones.

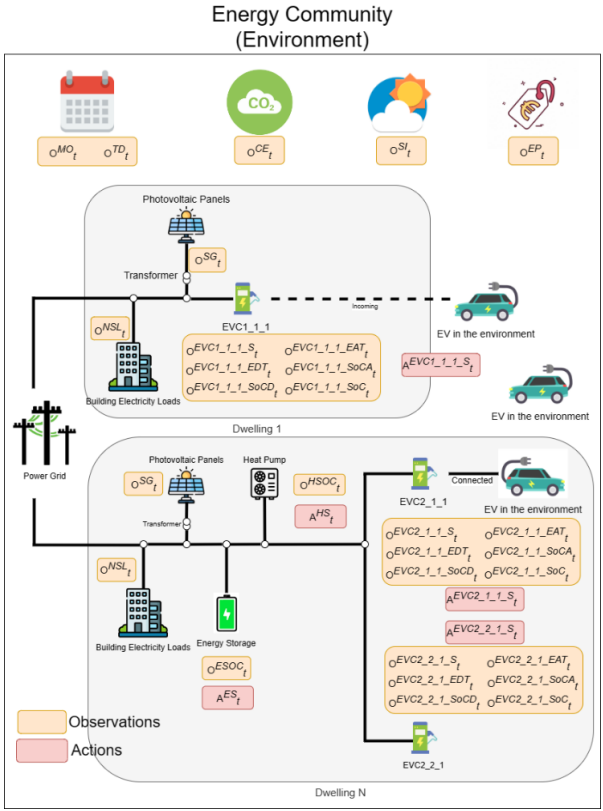

Figure 4 - Observations and Action in the CityLearn Framework

### D. Configuration File (Schema)

In the configuration file (the schema), each EV is delineated as a unique object, denoted by keys such as "EV_1", "EV_2", and so forth. Each EV object contains specific properties that define its characteristics. For instance, the *energy_simulation* property defines the .csv file used for energy consumption simulation (the pre-simulated dataset "electric_vehicle_id.csv"). The battery associated with each EV is defined as a nested object inside the EV, and it contains properties like capacity (the maximum energy that can be stored in the battery), and nominal_power (the power capacity of the battery).

The configuration files also offer the flexibility to customize characteristics for each building, as before, but now users can also add the chargers within them. Similarly to EVs, each building is outlined as a unique object, such as "Building_1". This object already contains properties to define the building's characteristics, including energy simulation, weather data, carbon intensity, pricing information and other energy models such as heat pumps and PVs. Chargers for EVs within the buildings can be added through the "chargers". Each charger is defined as a nested object within the building object, with properties such as the *nominal_power_charging, nominal_power_discharging, efficiency*. The name of the charger nested objset should follow the same convention as $EVC_{b\_n\_p}$. Efficiency curves can also be specified. This allows us to accurately represent the capabilities of the chargers in each building. This level of customization enables users to tailor their simulation scenarios to fit a wide range of urban setups and energy scenarios.

### V. EXPERIMENTAL SETUP

#### A. Simulation Scenario

In this work, a data fusion process was accomplished to construct two simulation environments compatible with the CityLearn platform, with one of the simulation scenarios also integrating the EVLearn module. This process ensured that all essential input for training and testing the EMSs was provided, offering environments that mirror, with the highest possible accuracy, real-world Energy Community/district energy management. Fig. 5 summarizes the components of the two simulation scenarios used during this work's case study.

Both simulation scenarios derive from a dataset [33] previously used within CityLearn Challenge [21]). It features 9 dwellings and 4 years of data. The dwellings are of different types, including a medium-sized office, a fast-food restaurant, a standalone retail store, a strip mall, and five medium-scale multifamily residences. It details air-to-water heat pumps, electric heaters, and on-site photovoltaic panels.

The original dataset used in the simulation scenarios contained detailed information on carbon emissions and weather patterns, data that is needed for EMSs. This data was further enriched by integrating real-time energy pricing from the Iberian wholesale energy market (OMIE)[1]. This addition, corresponding in length to the base dataset (i.e., four years of electricity prices for the 4-year dataset), aimed to enhance the real-world relevance of the simulation results.

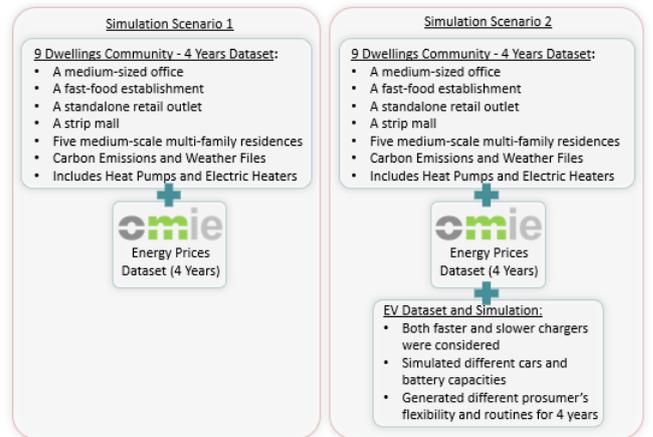

Figure 5 - Simulation Scenario Dataset Composition

---

[1] https://www.omie.es/en/market-results/daily/daily-market

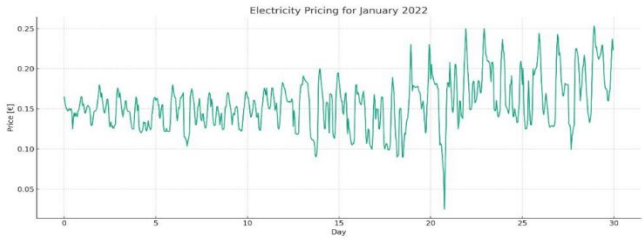

*Figure 6 – Real-Time Electricity Prices from OMIE for January 2022*

An example of a portion of these prices, January 2022 is illustrated in Fig. 7. The first scenario, SS1, examines the impact of not implementing EVLearn and not simulating EVs. This scenario only accounts for assets already at the CityLearn dataset. The second scenario, SS2, delves into V2G simulation, simulating EVs, EVCs and their energy flexibility as introduced during the course of this paper (Section III.E.). Refer to Appendix J of [34] for detail at each prosumer flexible assets.

*B. Test and evaluation methodology*

To evaluate the success of the objectives, a baseline and a set of Key Performance Indicators (KPIs) have been established according to the KPIs defined originally by [35].

KPIs are presented as a normalized value to each simulation scenario baseline. SS1 and SS2 baselines, respectively, are defined as the metrics when no control or optimization from an EMS is applied to the environment. In this context, the baseline represents the initial data observed and measured, indicating the EC consumption prior to the introduction of any intelligent management techniques controlling their flexible assets. The baseline applied for the EV extension of the simulation regards as baseline the consumption and normal charging behavior of EVs before any control is applied (i.e., the car reaches home and charges the maximum the charger can provide).

The algorithm chosen for the EMS is EnergAIze, proposed in [34] as a decentralized multi-agent reinforcement learning algorithm based on the MADDPG framework for optimizing and controlling energy assets within CityLearn [44]. The algorithm was trained for 15 episodes. Results are normalized to the no control baseline of each SS and are taken at the REC during the final episode as deterministic control.

## VI. DEMONSTRATION OF EVLEARN INTEGRATION

Figure 10 outputs the management results for a day using the EnergAIze algorithm for the presented simulation scenario 2 on V2G dynamics (Chapter V). The green shaded zones are when the vehicle was connected to the building's EVC. Let's break down the four plots in the figure:

- Building Electricity Consumption (Plot a): This graph shows the energy usage of a building before (in orange) and after (in blue) applying the MADDPG EMS. The energy usage prior to deploying the EMS (orange) provides a baseline benchmark, representing the scenario where no control or optimization measures have been applied.

- Charging Action Data (Plot b): depicts the energy $E_t^{EVC_{b\_n\_p}}$, in kWh, that was supplied by the EVC to the connected EV. This value, according to Eq. 1., was derived from the action value $a_{t-1}^{EVC_{b\_n\_p}}$ output as control action by the algorithm.

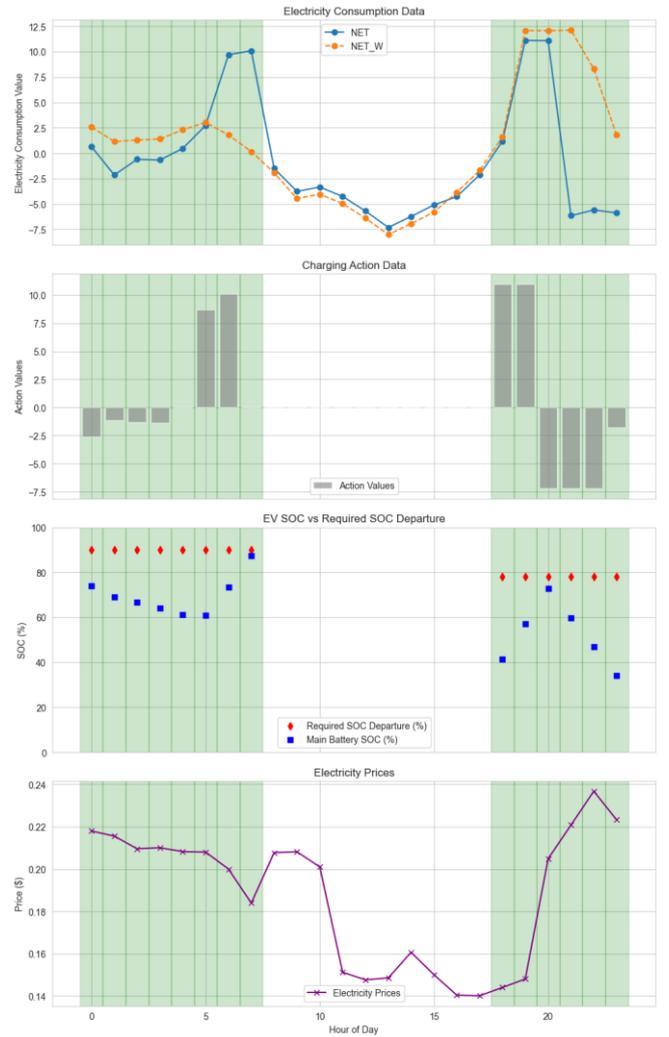

*Figure 7 – Energy Management during a day.*

- EV State of Charge (SoC) vs. Required SoC at Departure (Plot c): The blue squares show the actual battery charge level of the EV over time ($O_t^{EVC_{b,n,p}\_SoC}$), while the red dots indicate the desired battery charge level ($O_t^{EVC_{b,n,p}\_SoC\_D}$) needed when the EV is scheduled to leave. This comparison shows whether the smart system is effectively meeting the energy flexibility charging needs of prosumers.

- Electricity Prices (Plot d): The final graph illustrates the real-time cost of electricity throughout the day.

First, these preliminary results conclusively demonstrate the implementation and integration of the EVLearn simulation module within the CityLearn framework. From Fig. 10. we can conclude that the simulation was running with the complexities associated with EVs and EVCs, encompassing the pre-simulated routines of plug-in to and plug-out from required SoC at departure ($O_t^{EVC_{b,n,p}\_SoC\_D}$), articulating energy flexibility options in the form of the dynamics of charging and discharging (V2G). Moreover, plot b) and c) demonstrate how continuous action values and updates to the battery SoC were accomplished. These results highlight the EVs potential to enhance EC energy management.

In fact, some result analysis on energy management using EVs can be conducted from these preliminary results. For a complete result analysis refer to [34]. For example, refer to Fig. 10, plot c), at 7 am, and verify how the EV leaves with a SoC close to what was required by the prosumer. Also refer to Fig. 10 at 5am, or 6pm to 7pm and check how the algorithm learned to charge the battery at times of lower electricity prices, and the discharging happening from around 8pm onwards, when prices increase, and the dwelling consumption is higher (peak reduction and cost reduction).

From plot a) it is possible to check how the late-night peak in consumption (from 6pm to 10 pm) was shifted towards the morning, through periods of lower energy prices (and possibly more renewable energy production, mainly solar). Also check on how this building outputs close to 5kWh between 9 to 11 pm, to help other Energy Community members, which do not have batteries, to consume less energy from the grid during the night, and preferably consuming from the neighbors at the set lower discounted price.

Evolving from the more detailed analysis of one energy management strategy, Table 3 provides a broader, community/district wide comparison of KPIs for the dataset and two simulation scenarios presented in Chapter V.

When observing the results at the broader EC/district level, the EMS's KPIs at scenario 2 standout. The total carbon emissions (G) and energy cost were cut by close to 12% compared to the baseline. Zero net energy (Z) also recorded an enhancement, marked by a 6% increase in self-consumed energy against baseline numbers. Of particular interest is the average ramping metric (R), the daily peak average (P) and the 1-Load Factor (1-L), where the Marlisa within the EV environment exhibited a reduction of 35%, 21% and 13% respectively, when compared to the baseline. When compared to the no EV scenario. Scenario 1 shows smaller gains when compared to scenario 2, which can be attributed to the greater degree of energy flexibility an EV can offer to an EMS, when compared to a simpler Heat Pump.

*Table 3 - Community-Level KPIs*

| KPI | SS 1 | SS 2 |
|---|---|---|
| Electricity Consumption (D) | -3.61% | -12.46% |
| Electricity Price (C) | -4.81% | -11.35% |
| Carbon Emissions (G) | -6.14% | -11.84% |
| Zero Net Energy(Z) | 2.94% | 6.22% |
| Average Daily Peak (P) | -15.74% | -20.80% |
| Ramping (R) | -13.49% | -35.22% |
| 1 - Load Factor (1-L) | -7.16% | -13.43% |

Of special notice is the reduction obtained in ramping, as the EV charging can be smoothed without impacting the prosumers' comfort (e.g., by charging during the night). This difference from SS 1 to SS 2 underscores the impact of EVs in energy management strategies and underlines the value this paper brings by incorporating EV simulation into the broader scope of other flexible energy assets.

## VII. CONCLUSIONS

This paper addressed the pressing challenge of integrating EVs and energy management strategies such as V2G and G2V within EC EMSs, a step towards mitigating the impacts of climate change. By developing the EVLearn simulation module and integrating it into the established CityLearn framework, we have provided an advancement to a recognized gap in the simulation of energy management strategies.

The modelling and integration of EVs, their underlying EVCs and associated energy flexibility, not only enables a more comprehensive and integrated simulation of urban energy systems but also facilitates the exploration and optimization of energy management strategies that incorporate the dynamic and potentially transformative role of EVs in enhancing grid stability and efficiency.

The presented results of a comparative analysis with and without the integration of EVs into EMSs, validates the EVLearn development and integration into CityLearn. Moreover, the preliminary results underscore the substantial benefits that EvLearn can offer by paving the way for innovative advancements in the field by enabling the exploration and analysis of straightforward G2V and more intricate V2G solutions, such as the one presented in this work, ultimately fostering the seamless integration of EVs.

To further enhance the capabilities of the simulation and its real-world applicability, future work can focus on several improvements. One area for improvement involves refining the simulation's time step granularity, transitioning from the current one-hour intervals to shorter periods, such as five minutes or less. Additionally, incorporating simulation models of other energy-flexible physical assets, such as washing machines, into the framework, alongside refining existing models to encapsulate greater complexity, would broaden the evaluative scope of algorithms and enrich case studies available for research.

The deployment of advanced visualization tools within the framework is another critical step. Regarding specifically to the extension module, EVLearn, future work might delve into integrating newer, more insightful, EV energy flexibility datasets. Such enhancements are instrumental in narrowing the divide between theoretical models and their practical applications, setting the stage for the development of more sophisticated and comprehensive EMSs.

ACKNOWLEDGMENT

This paper is supported by the OPEVA project that has received funding within the Chips Joint Undertaking (Chips JU) from the European Union's Horizon Europe Programme and the National Authorities (France, Czechia, Italy, Portugal, Turkey, Switzerland), under grant agreement 101097267. The paper is also supported by Arrowhead PVN, proposal number 101097257. Views and opinions expressed are however those of the author(s) only and do not necessarily reflect those of the European Union or Chips JU. Neither the European Union nor the granting authority can be held responsible for them.

REFERENCES

[1] G. A. Pagani and M. Aiello, "The Power Grid as a complex network: A survey," *Physica A: Statistical Mechanics and its Applications*, vol. 392, no. 11, pp. 2688–2700, Jun. 2013, doi: 10.1016/J.PHYSA.2013.01.023.

[2] M. Victoria *et al.*, "Solar photovoltaics is ready to power a sustainable future," *Joule*, vol. 5, pp. 1041–1056, 2021, doi: 10.1016/j.joule.2021.03.005.

[3] K. Bódis, I. Kougias, A. Jäger-Waldau, N. Taylor, and S. Szabó, "A high-resolution geospatial assessment of the rooftop solar photovoltaic potential in the European Union," *Renewable and Sustainable Energy Reviews*, vol. 114, p. 109309, Oct. 2019, doi: 10.1016/J.RSER.2019.109309.

[4] K. Surana and S. M. Jordaan, "The climate mitigation opportunity behind global power transmission and distribution," *Nature Climate Change 2019 9:9*, vol. 9, no. 9, pp. 660–665, Aug. 2019, doi: 10.1038/s41558-019-0544-3.

[5] M. Sheha, K. Mohammadi, and K. Powell, "Solving the duck curve in a smart grid environment using a non-cooperative game theory and dynamic pricing profiles," Energy Convers Manag, vol. 220, p. 113102, Sep. 2020, doi: 10.1016/J.ENCONMAN.2020.113102.

[6] J. Zhong, M. Bollen, and S. Rönnberg, "Towards a 100% renewable energy electricity generation system in Sweden," *Renew Energy*, vol. 171, pp. 812–824, Jun. 2021, doi: 10.1016/J.RENENE.2021.02.153.

[7] G. Dileep, "A survey on smart grid technologies and applications," *Renew Energy*, vol. 146, pp. 2589–2625, Feb. 2020, doi: 10.1016/J.RENENE.2019.08.092.

[8] I. Antonopoulos *et al.*, "Artificial intelligence and machine learning approaches to energy demand-side response: A systematic review," *Renewable and Sustainable Energy Reviews*, vol. 130, p. 109899, Sep. 2020, doi: 10.1016/J.RSER.2020.109899.

[9] C. Bergaentzlé, I. G. Jensen, K. Skytte, and O. J. Olsen, "Electricity grid tariffs as a tool for flexible energy systems: A Danish case study," *Energy Policy*, vol. 126, pp. 12–21, Mar. 2019, doi: 10.1016/J.ENPOL.2018.11.021.

[10] T. Fonseca, L. L. Ferreira, J. Landeck, L. Klein, P. Sousa, and F. Ahmed, "Flexible Loads Scheduling Algorithms for Renewable Energy Communities," *Energies 2022, Vol. 15, Page 8875*, vol. 15, no. 23, p. 8875, Nov. 2022, doi: 10.3390/EN15238875.

[11] T. Fonseca, L. L. Ferreira, L. Klein, J. Landeck, and P. Sousa, "Flexigy Smart-grid Architecture", doi: 10.5220/0010918400003118.

[12] [13] S. S. Ravi and M. Aziz, "Utilization of Electric Vehicles for Vehicle-to-Grid Services: Progress and Perspectives," *Energies 2022, Vol. 15, Page 589*, vol. 15, no. 2, p. 589, Jan. 2022, doi: 10.3390/EN15020589.

[13] A. Hoekstra, "The Underestimated Potential of Battery Electric Vehicles to Reduce Emissions," *Joule*, vol. 3, no. 6, pp. 1412–1414, Jun. 2019, doi: 10.1016/J.JOULE.2019.06.002.

[14] M. Muratori, "Impact of uncoordinated plug-in electric vehicle charging on residential power demand - supplementary data," Jun. 2017, doi: 10.7799/1363870.

[15] T. A. Skouras, P. K. Gkonis, C. N. Ilias, P. T. Trakadas, E. G. Tsampasis, and T. V. Zahariadis, "Electrical Vehicles: Current State of the Art, Future Challenges, and Perspectives," *Clean Technologies 2020, Vol. 2, Pages 1-16*, vol. 2, no. 1, pp. 1–16, Dec. 2019, doi: 10.3390/CLEANTECHNOL2010001.

[16] F. Gonzalez Venegas, M. Petit, and Y. Perez, "Active integration of electric vehicles into distribution grids: barriers and frameworks for flexibility services," 2021.

[17] D. Qiu, Y. Wang, W. Hua, and G. Strbac, "Reinforcement learning for electric vehicle applications in power systems:A critical review," *Renewable and Sustainable Energy Reviews*, vol. 173, p. 113052, Mar. 2023, doi: 10.1016/J.RSER.2022.113052.

[18] K. Nweye, B. Liu, P. Stone, and Z. Nagy, "Real-world challenges for multi-agent reinforcement learning in grid-interactive buildings," *Energy and AI*, vol. 10, p. 100202, Nov. 2022, doi: 10.1016/J.EGYAI.2022.100202.

[19] J. R. Vázquez-Canteli and Z. Nagy, "Reinforcement learning for demand response: A review of algorithms and modeling techniques," *Appl Energy*, vol. 235, pp. 1072–1089, Feb. 2019, doi: 10.1016/J.APENERGY.2018.11.002.

[20] J. R. Vazquez-Canteli, S. Dey, G. Henze, and Z. Nagy, "CityLearn: Standardizing Research in Multi-Agent Reinforcement Learning for Demand Response and Urban Energy Management," Dec. 2020, doi:10.48550/arXiv.2012.10504

[21] J. R. Vázquez-Canteli, J. Kämpf, G. Henze, and Z. Nagy, "CityLearn v1.0: An OpenAI gym environment for demand response with deep reinforcement learning," *BuildSys 2019 - Proceedings of the 6th ACM International Conference on Systems for Energy-Efficient Buildings, Cities, and Transportation*, pp. 356–357, Nov. 2019, doi: 10.1145/3360322.3360998.

[22] P. Scharnhorst *et al.*, "Energym: A Building Model Library for Controller Benchmarking," *Applied Sciences 2021, Vol. 11, Page 3518*, vol. 11, no. 8, p. 3518, Apr. 2021, doi: 10.3390/APP11083518.

[23] D. Blum *et al.*, "Building optimization testing framework (BOPTEST) for simulation-based benchmarking of control strategies in buildings," *J Build Perform Simul*, vol. 14, no. 5, pp. 586–610, Sep. 2021, doi: 10.1080/19401493.2021.1986574.

[24] D. B. Crawley et al., "EnergyPlus: creating a new-generation building energy simulation program," Energy Build, vol. 33, no. 4, pp. 319–331, Apr. 2001, doi: 10.1016/S0378-7788(00)00114-6.

[25] P. A. Fritzson, "Principles of object-oriented modeling and simulation with Modelica 2.1," p. 897, 2004.

[26] N. Blair et al., "System Advisor Model, SAM 2014.1.14: General Description," NREL Report No. TP-6A20-61019, National Renewable Energy Laboratory, Golden, CO, no. February, p. 13, Feb. 2014, doi: 10.2172/1126294.

[27] R. Buyya and M. Murshed, "GridSim: A Toolkit for the Modeling and Simulation of Distributed Resource Management and Scheduling for Grid Computing".

[28] L. Thurner *et al.*, "pandapower - an Open Source Python Tool for Convenient Modeling, Analysis and Optimization of Electric Power Systems," *IEEE Transactions on Power Systems*, vol. 33, no. 6, pp. 6510–6521, Sep. 2017, doi: 10.1109/TPWRS.2018.2829021.

[29] T. B. Pedersen, L. Siksnys, and B. Neupane, "Modeling and Managing Energy Flexibility Using FlexOffers," *2018 IEEE International Conference on Communications, Control, and Computing Technologies for Smart Grids, SmartGridComm 2018*, Dec. 2018, doi: 10.1109/SMARTGRIDCOMM.2018.8587605.

[30] L. Calearo, M. Marinelli, and C. Ziras, "A review of data sources for electric vehicle integration studies," *Renewable and Sustainable Energy Reviews*, vol. 151, p. 111518, Nov. 2021, doi: 10.1016/J.RSER.2021.111518.

[31] Y. Amara-Ouali, Y. Goude, P. Massart, J. M. Poggi, and H. Yan, "A Review of Electric Vehicle Load Open Data and Models," *Energies 2021, Vol. 14, Page 2233*, vol. 14, no. 8, p. 2233, Apr. 2021, doi: 10.3390/EN14082233.

[32] G. Brockman *et al.*, "OpenAI Gym," Jun. 2016, doi:10.48550/arXiv.1606.01540

[33] Z. Nagy, J. R. Vázquez-Canteli, S. Dey, and G. Henze, "The citylearn challenge 2021," in *Proceedings of the 8th ACM International Conference on Systems for Energy-Efficient Buildings, Cities, and Transportation*, New York, NY, USA: ACM, Nov. 2021, pp. 218–219. doi: 10.1145/3486611.3492226.

[34] T. C. C. Fonseca, "A Multi-Agent Reinforcement Learning Approach to Integrate Flexible Assets into Energy Communities," Oct. 2026, Accessed: Mar. 27, 2024. https://recipp.ipp.pt/handle/10400.22/24068

[35] J. R. Vazquez-Canteli, G. Henze, and Z. Nagy, "MARLISA: Multi-Agent Reinforcement Learning with Iterative Sequential Action Selection for Load Shaping of Grid-Interactive Connected Buildings," *BuildSys 2020 - Proceedings of the 7th ACM International Conference on Systems for Energy-Efficient Buildings, Cities, and Transportation*, pp. 170–179, Nov. 2020, doi: 10.1145/3408308.3427604.